\documentclass[twocolumn,aps,prc,floatfix]
{revtex4}

\usepackage[dvips]{graphicx}         %
\usepackage{bm}                      
\usepackage{mathptmx}                
\usepackage{dcolumn}                 
\usepackage{multirow}                %

\def\bea{\begin{eqnarray}} \def\eea{\end{eqnarray}}
\def\beq{\begin{equation}} \def\eeq{\end{equation}}
\def\bal#1\eal{\begin{align}#1\end{align}}
\def\bse#1\ese{\begin{subequations}#1\end{subequations}}

\begin{document}

\title{Interplay of effects of neutron skins in coordinate space and proton skins in momentum space on hard photons emission in heavy-ion collisions around Fermi Energy }

\author{Wen-Mei Guo$^1$\footnote{guowenmei@ahnu.edu.cn}}
\author{Bao-An Li$^2$\footnote{Bao-An.Li@TAMUC.edu}}
\author{Gao-Chan Yong$^3$\footnote{yonggaochan@impcas.ac.cn}}
\affiliation{
$^1${Department of Physics, Anhui Normal University, Wuhu 241000, China}\\
$^2${Department of Physics and Astronomy, Texas A\&M University-Commerce, Commerce, Texas 75429-3011, USA}\\
$^3${Institute of Modern Physics, Chinese Academy of Sciences, Lanzhou 730000, China}}

\date{\today}

\begin{abstract}
Within an isospin- and momentum-dependent Boltzmann-Uehling-Uhlenbeck (IBUU) transport model, we investigate the hard photons emission from neutron-proton bremsstrahlung in
reaction system of $^{208}Pb+^{208}Pb$ around Fermi energy.
Effects of neutron skins in coordinate ($r$) space and proton skins in momentum ($k$) space on the time evolution, the angular distribution, and the transverse momentum spectra of hard photons with different energies are studied. It is shown that the emission of direct hard photons is sensitive to the neutron skin, which has larger effects for more energetic hard photons. Meanwhile, we find that the proton skins have an important influence on the emission of direct hard photons, and its effect is actually even larger than that of neutron skins. It needs to take the effect of proton skins into account when we determine the size of neutron skins by comparing transport mode predictions of hard photons with the corresponding experiment measurements.

\end{abstract}



\maketitle

\section{Introduction}
For two decades, the neutron skin thickness of nuclei as a sensitive probe of the nuclear symmetry energy has attracted much attention in nuclear physics and astrophysics \cite{Brown00,Brendan21,Cente09,Roca11}. It is defined as the difference between the root-mean-squared (rms) radii of neutrons and protons, i.e., $\Delta r_{np}=<r^2_n>^{1/2}-<r^2_p>^{1/2}$. Much effort has been devoted to measuring the size of neutron skin experimentally, e.g., photopion production, pionic and antiprotonic atoms method at CERN \cite{Trzc01,Jastr04,Horo14}, parity-violating electron scattering at the Jefferson Lab (PREX-I and PREX-II experiment) \cite{PREXI12,PREXII21}.
Theoretically, lots of theories, e.g., mean-field models \cite{Roca11,Chen10}, \emph{ab initio} computations \cite{Rein21,Hubs22} and
droplet models \cite{Cent09,Daniel03}, have been used to study the neutron skin thicknesses of neutron-rich nuclei.
Moreover, extractions of $\Delta r_{np}$ in $^{208}Pb$ and $^{209}Bi$ from different hadronic probes have been made \cite{Zeni10,Klos07}, although there might be some degree of model dependence involved. However, the community has not reach a consensus on precise values of neutron skins of heavy nuclei. Therefore, it is of great importance to improve the accuracy of the extraction of neutron skin thickness for helping us to explore the density dependence of nuclear symmetry energy, which can conduce to understand properties of neutron-rich matter and even of neutron star for us.

It is well known that observables of heavy-ion collisions are sensitive to the
initial phase space distributions of nucleons or quarks and gluons in the colliding nuclei. For example, it was proposed that the yield ratios of neutron-proton and that of $^3H-^3He$ could serve as the sensitive probes to neutron skin thickness in heavy-ion collisions, see, e.g., Refs. \cite{Sun10,Yan19} for recent reviews. The neutron skin effect also has been studied at both intermediate \cite{GFWei} and relativistic heavy-ion collisions \cite{Lihl20,Hamme20,Giaca-new}. It is also well known that hard photons are promising probes to study the reaction dynamics in nuclear physics \cite{Yong08}, because they only interact weakly with the nuclear medium through electromagnetic force unlike hadronic probes inevitably suffering from distortions due to strong interactions in the final stage. For hard photons,
usually defined as $\gamma$-ray spectra above 30 MeV (distinct from the giant dipole
resonance emission \cite{Giul06}), experimental and theoretical studies consistently indicate
that they are emitted mainly in incoherent proton-neutron bremsstrahlung,
$p+n\rightarrow p+n+\gamma$, during the early stage of heavy-ion collisions
\cite{Gan94,Biro87,Bauer86,Liu08,Yong17}. This part of hard photons associated with
the first-chance proton-neutron collisions are called as direct photons.
Recently, Refs. \cite{Wei15,Wang22} have put forward that hard photon emission can be taken as an experimental observable to extract information on neutron skin thickness. On the other hand, nucleon-nucleon short-range correlations (SRCs) due to the tensor components and/or the repulsive core of nuclear forces have attracted much attention in recent years \cite{Arri12,Ciofi15,Hen:2016kwk}. The SRC will
lead to the formation of a high-momentum tail (HMT) in the single-nucleon momentum distribution. Incorporating the SRC effects in extended Thomas-Fermi model, Ref.~\cite{Cai16} shows that proton skins in $k$ space coexist with neutron skins in $r$ space in heavy nuclei and their correlation is governed by Liouville's theorem and Heisenberg's uncertainty principle. Investigating SRC effects in nuclei, nuclear reactions and neutron stars is a major task in nuclear physics \cite{Sorensen}.

In the present work, the Thomas-Fermi approximation to the nucleon kinetic-energy density, which incorporates the SRC effects,
was considered in the IBUU transport model. The interplay of effects of neutron skins in $r$ space and proton skins in $k$ space on hard photons emission in heavy-ion collisions around Fermi Energy was studied.

\section{Theoretical framework}
In this study, we adopt the updated IBUU transport model which is originated from
the IBUU04 model \cite{LiBA04,Liba05}.
The initial density distributions of nucleons in projectile and target are obtained from the two-parameter Fermi (2pF) distribution widely used in the literature, i.e.,
\beq
\rho_J(r)=\rho^J_0\{1+exp[(r-c_J)/a_J]\}^{-1},
\eeq
where $c_J$ and $a_J$ are the half-density radius and diffuseness parameter, respectively.

With the fact that the momentum distribution of nucleons has a high momentum tail due to SRCs, the parameterized nucleon momentum distribution according to Ref.~\cite{Cai16,Ohen15} is taken into account in our IBUU model and given as follows:
\beq
n^J_k(\rho,\delta)=\left\{
\begin{array}{ll}
\Delta_J, &0< |\textbf{k}|< k^J_F,\\ \
\\C_J(k^J_F/|\textbf{k}|^4), &k^J_F<|\textbf{k}|<\phi_J k^J_F,
\end{array}
\right.
\eeq
where $k^J_F$ is the Fermi momentum of the nucleon J. The $\Delta_J$ denotes the depletion of the Fermi sea with respect to the
step function for a free Fermi gas. The parameters $\Delta_J$, $C_J$ and $\phi_J$ depend linearly on the isospin asymmetry $\delta=(\rho_n-\rho_p)/\rho$ in a general form of $Y_J=Y_0(1+Y_1\tau^J_3\delta)$ where $\tau^n_3=+1$ and $\tau^p_3=-1$\cite{Hass04,Rios14,Lizh16}.
The amplitude $C_J$ and the high-momentum cutoff coefficient $\phi_J$ determine the fraction of
nucleons in the HMT via $x^{HMT}_J=3C_J(1-\phi^{-1}_J)$. The above parameters are constrained by the the normalization condition $[2/(2\pi)^3]\int^{\infty}_0n^J_{\textbf k}(\rho,\delta)d{\textbf k}=(k^J_F)^3/3\pi^2$, the EOS of pure neutron matter (PNM) from microscopic many-body theories \cite{Schw05,Epel09,Gezer10,Stew10,Tews13,Gezer13} and the systematic analysis of many experiments about the percentage of nucleons in the HMT on the symmetry nuclear matter (SNM) \cite{Ohen15,Ohen155,Ohen14}. According to Ref.~\cite{Cai16}, the HMT-exp parameters sets of $x^{HMT}_{SNM}=28\%$, $x^{HMT}_{PNM}=4\%$, which caused $C_0=0.161$, $C_1=-0.25$, $\phi_0=2.38$ and $\phi_1=-0.56$, were adopted in our study. With this nucleon momentum distribution, the proton skin was displayed in $k$ space, and its thickness grows with the isospin asymmetry \cite{Cai16}. Effects of the SRC/HMT described above on nuclear symmetry energy and properties of neutron stars have been studied extensively. For a review, see, e.g., Ref.~\cite{Li18}.

Incorporating the SRC effects in the extended Thomas-Fermi approximation, the nucleon kinetic-energy density profile in finite nuclei is modified as bellow \cite{Cai16},
\beq
\varepsilon^{kin}_J(r)=\frac{1}{2M}[\alpha^{\infty}_J\rho^{5/3}_J(r)\Phi_J
+\frac{\eta_J}{36}\frac{[\nabla\rho_J(r)]^2}{\rho_J(r)}+\frac{1}{3}\Delta\rho_J(r)].
\eeq
The first term with $\alpha^{\infty}_J=(3/5)(3\pi^2)^{2/3}$ is the bulk part as if nucleons are in infinite nuclear matter, in which
$\Phi_J=1+C_J(5\phi_J+3/\phi_J-8)>1$ is determined by properties of the HMT. It makes the bulk part of the kinetic-energy density
enhanced more for protons than neutrons in neutron-rich systems, since relatively more protons are depleted from the Fermi sea to form a proton skin in the HMT. Here, $\Phi_p=2.09$ and $\Phi_n=1.60$ was used for isospin asymmetry $\delta=0.21$. The second term originally proposed by Weizs\"{a}cker \cite{Weizs35} is called the surface term in this study as it is very sensitive to surface properties of finite nuclei. Its strength factor $\eta_{J}$ has been under debate and was found to affect significantly the halo and skin nature of the surfaces of heavy nuclei \cite{Cai16,Brack85,Lucy2016}. It can be constrained using the experimental information of the density profile and the average kinetic-energy $<E_J^{kin}>$ of nucleons. The third term relating to Laplacian operator is very small because of the smooth nuclear surface.

We adopt the same formalism from Ref.~\cite{Cai16} for the proton skin in $k$ space, which is similar to the measure of the neutron skin in $r$ space ($\Delta r_{np}=<r^2_n>^{1/2}-<r^2_p>^{1/2}$ with $<r^2_{n/p}>^{1/2}$ the rms radius of neutrons or protons), that is the difference between the average kinetic energies of protons and neutrons, i.e., $\Delta E^{kin}_{pn}\equiv<E^{kin}_p>-<E^{kin}_n>$ with
\beq
<E^{kin}_J>=\int^{\infty}_0\varepsilon^{kin}_J(r)d\textbf{r}/\int^{\infty}_0\rho_J(r)d\textbf{r}\equiv<k^2_J>/2M,
\eeq

In this study, the following isospin- and momentum- dependent mean field single-nucleon potential
is used \cite{Das03},
\begin{eqnarray}
U(\rho,\delta,\textbf{p},\tau)&=&A_u(x)\frac{\rho_{\tau'}}{\rho_0}+A_l(x)\frac{\rho_{\tau}}{\rho_0}\nonumber\\
& &+B(\frac{\rho}{\rho_0})^{\sigma}(1-x\delta^2)
   -8x\tau\frac{B}{\sigma+1}\frac{\rho^{\sigma-1}}{\rho_0^\sigma}\delta\rho_{\tau'}\nonumber\\
& &+\frac{2C_{\tau,\tau}}{\rho_0}\int
d^3\,\textbf{p}'\frac{f_\tau(\textbf{r},\textbf{p}')}{1+(\textbf{p}-\textbf{p}')^2/\Lambda^2}\nonumber\\
& &+\frac{2C_{\tau,\tau'}}{\rho_0}\int
d^3\,\textbf{p}'\frac{f_{\tau'}(\textbf{r},\textbf{p}')}{1+(\textbf{p}-\textbf{p}')^2/\Lambda^2},
\end{eqnarray}
where $\tau=1/2\ (-1/2)$ for neutrons (protons),
$\delta=(\rho_n-\rho_p)/(\rho_n+\rho_p)$ is the isospin asymmetry,
and $\rho_n$, $\rho_p$ denote neutron and proton densities,
respectively. Specifically, the parameters $A_u(x)$, $A_l(x)$, $B$,
$C_{\tau,\tau}$, $C_{\tau,\tau'}$ $\sigma$, and $\Lambda$ with SRCs are updated
by fitting empirical nature of nuclear matter, for example,
the saturation density $\rho_0=0.16$ fm$^{-3}$, the binding energy $E_0=-16$ MeV,
the incompressibility $K_0=230$ MeV and the isoscalar effective mass
$m^*_s=0.7m$ and so on. $f_{\tau}(\textbf{r},\textbf{p})$ is
the phase-space distribution function at coordinate \textbf{r} and
momentum \textbf{p}. Different $x$ parameters can be used to mimic
different forms of the symmetry energy predicted by various many-body
theories without changing any property of the symmetric nuclear matter
and the symmetry energy at normal density. And for nucleon-nucleon
scattering, the isospin-dependent reduced in-medium nucleon-nucleon
cross section is adopted.

Hard photons production in heavy-ion reactions at intermediate energy has been
studied theoretically and experimentally in a number of works
\cite{Yong08,Bauer86,Cass90,Yong171,Yong18,Schu97,Bertsch88}.
Although the elementary cross section for the $p+n\rightarrow p+n+\gamma$ process
is still model dependent \cite{Gan94,Naka86,Sch91,Timm06},
the calculations from theoretical reaction models are
able to reasonably reproduce all qualitative features of the hard photon experimental
data \cite{Cass90}. In this work, we adopt the following approach for the probability of hard photons
production based on the neutral scalar $\sigma$ meson exchange model by
standard quantum field theory \cite{Gan94}, in which more quantum mechanical
effects was considered. It is fitted by an analytical expression
\beq
\frac{d^2p_{\gamma}}{d\Omega dE_{\gamma}}=1.671\times10^{-7}\frac{(1-y^2)^{\alpha}}{y},
\eeq
where $y=E_{\gamma}/E_{max}$, $\alpha=0.7319-0.5898\beta_i$, $E_{\gamma}$ is energy of emitted photon,
$E_{max}$ is the total energy available in the proton-neutron c.m. system, $\beta_i$
is the nucleon initial velocity. In the program, the effects of Pauli blocking in final
state at the $p+n \rightarrow p+n+\gamma$ process are also taken into account \cite{Bauer86}.
The emissions of photons here are assumed to be isotropic in the
proton-neutron c.m. frame, therefore one obtains the single differential elementary probability
of photon by averaging the solid angle over $4\pi$, i.e.,
\beq
p_{\gamma}=\frac{dN}{dE_{\gamma}}=2.1\times10^{-6}\frac{(1-y^2)^{\alpha}}{y}.
\eeq

\section{Results and Discussions}

\begin{figure}[t!]
\centering
\setlength{\abovecaptionskip}{-4.cm}
\includegraphics[width=0.9\textwidth]{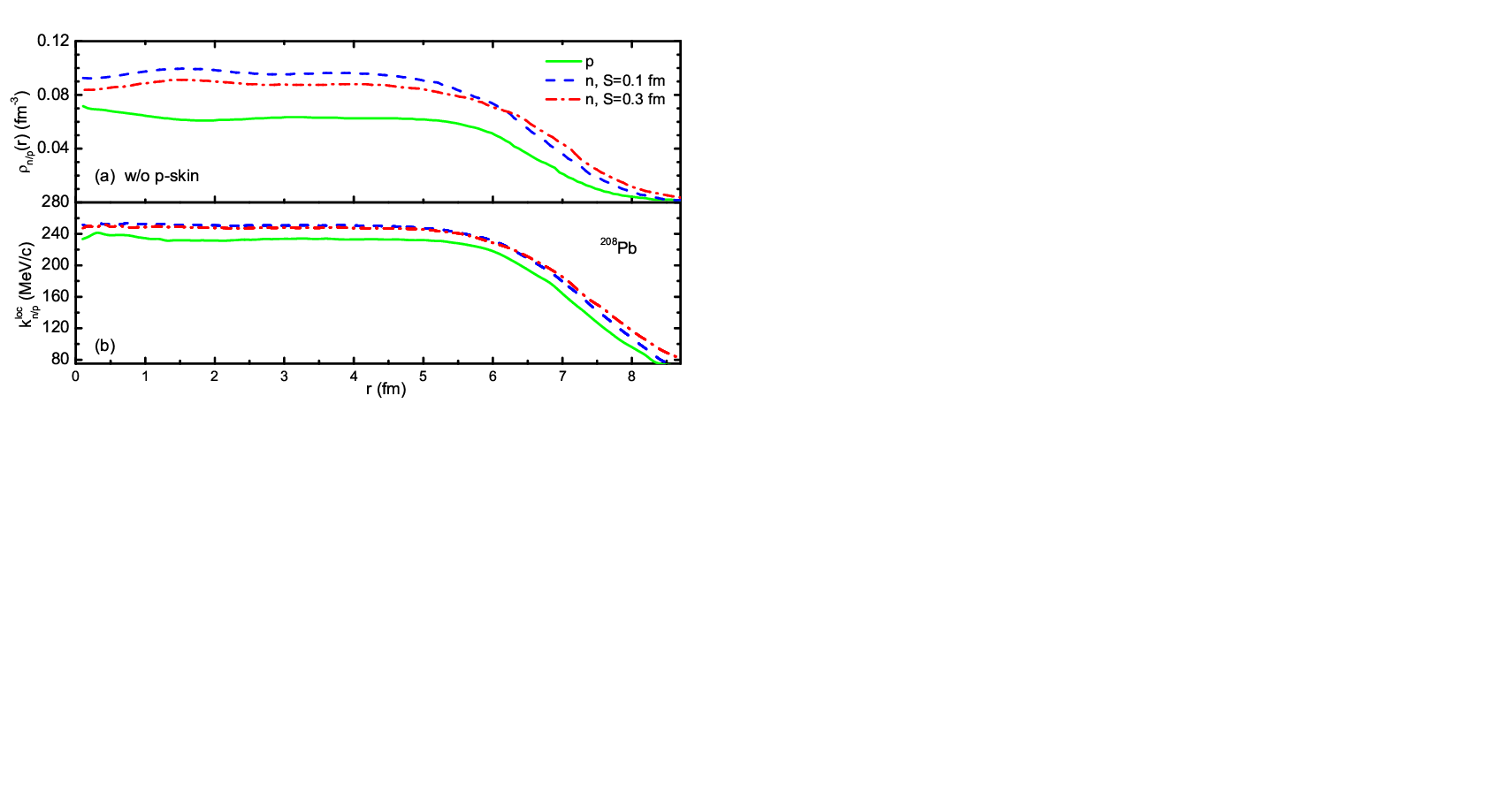}
\caption{Input local density (a) and calculated momentum profiles (b) for nucleons in $^{208}Pb$ and for neutrons with neutron skin thicknesses $S=0.1$ fm and $S=0.3$ fm. In this case the nucleon kinetic-energy density of colliding nuclei just includes the bulk part.} \label{dens-k1}
\end{figure}

As a comparison, we calculate the initial density distribution of nucleons in colliding nuclei from the Skyrme-Hartree-Fork calculations with the Skyrme $M^*$ parameter set \cite{Frie86} used in the IBUU model before. And the parameterized nucleon momentum distribution which includes the HMT mentioned above has been adopted. Shown in Fig.~1 is the nucleons' average local density (a) and momentum (b) as a function of radius $r$ in $^{208}Pb$. The solid line represents protons. The dash line and dash dot line are for neutrons with neutron skin thicknesses $S=0.1$ fm and $S=0.3$ fm, respectively. The momentum profiles in Fig.~1(b) are calculated from Eq.(4) in which the nucleon kinetic-energy density just includes the bulk part. We can see that neutrons have higher local momenta due to their higher densities than protons in the entire radius area of $^{208}Pb$. However, the protons have higher local momenta than neutrons in the surface area as shown in Fig.~2(b), when adopting the extended Thomas-Fermi approximation in the nucleon kinetic-energy density in which the SRC effects also were incorporated. The reason is that protons have larger values of the Weizs\"{a}cker surface term $(\nabla\rho_J/\rho_J)^2$. To evaluate the surface term, we adopt here the specified nucleon's density profile Eq.(1) by adjusting the diffuseness parameter of neutron density to obtain different neutron skin thicknesses for $^{208}Pb$ as shown in Fig.~2(a), in which the profile and numerical value are almost the same with that shown in Fig.~1(a). Therefore, according to the Eq.~(3), we have approximately in the surface area $k_{p/n}(r)\approx 1/(72M)(\nabla\rho_{p/n}/\rho_{p/n})^2\approx 1/(72Ma_{p/n}^2)$, leading to $k_p^{loc}(r)>k_n^{loc}(r)$ since the protons' surface diffuseness $a_p$ for protons is normally much less than the $a_n$ for neutrons in heavy nuclei. From Fig.~2(a) and (b), more interestingly, we can see that the coexistence of a proton skin in $k$ space and a neutron skin in $r$ space reappeared as shown in Ref.~\cite{Cai16}. In the current extended Thomas-Fermi approximation, the SRC effects were mainly included in the first term of the nucleon kinetic-energy density. It makes the local momenta of protons increase slightly even close to that of neutrons in interior area of nucleus.
\begin{figure}[t!]
\centering
\setlength{\abovecaptionskip}{-4.cm}
\includegraphics[width=0.9\textwidth]{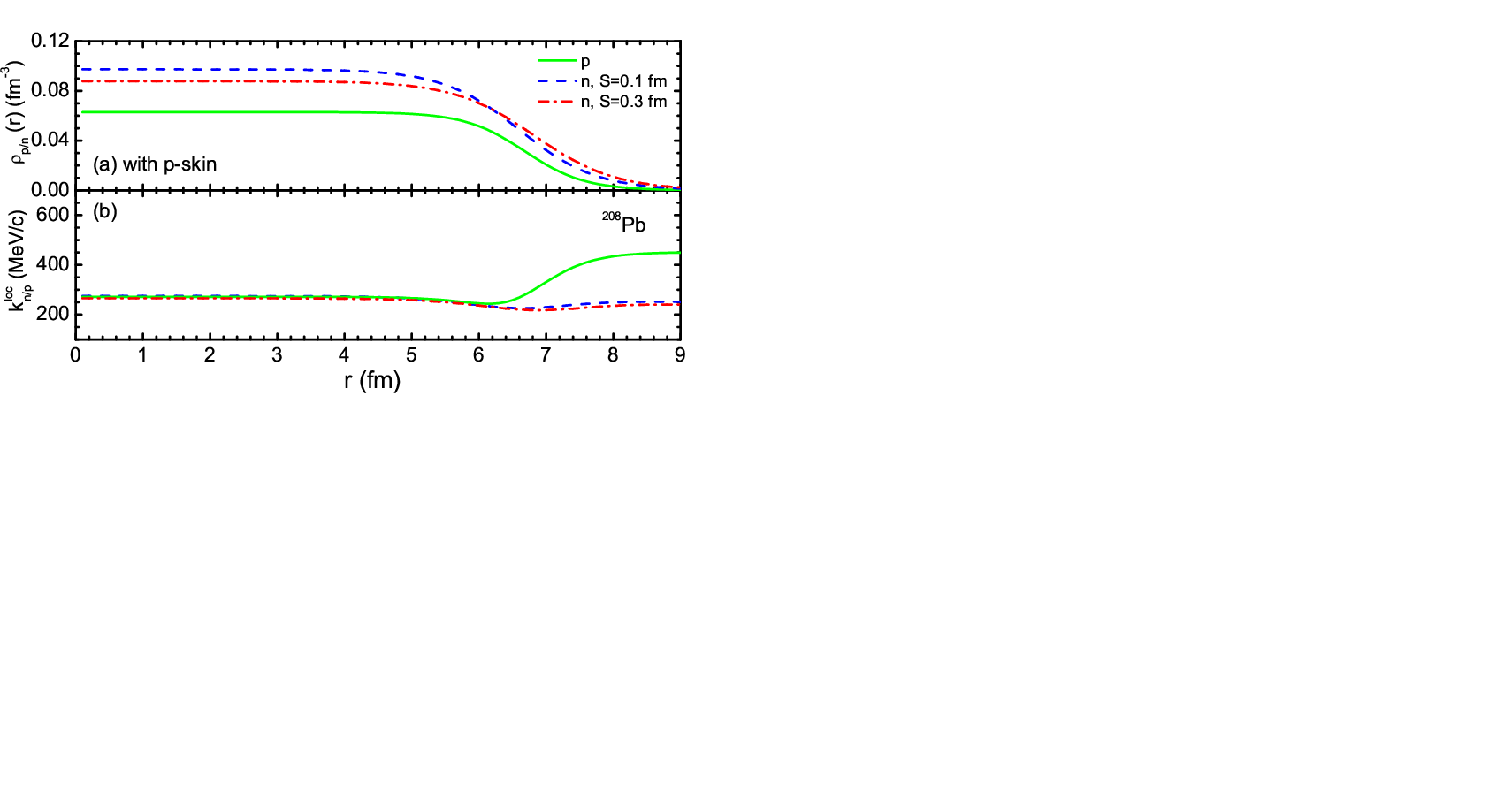}
\caption{Same with Fig.~1, but considering the surface term in the nucleon kinetic-energy density of colliding nuclei.} \label{dens-k2}
\end{figure}

In order to avoid the hadronic probes, while valuable, interact with the nuclear medium that hinder the clean determination of neutron skin, the hard photons as a promising probe for neutron skin are studied. Figure 3 shows effects of proton skin in $k$ space and neutron skin in $r$ space on the time evolutions of hard photons with energies of $E_{\gamma}=100,\ 150,\ 200$ and 250 MeV respectively in peripheral collisions $^{208}Pb+^{208}Pb$ at $E_b=45$ MeV/nucleon with impact parameter $b=11$ fm. It shows that the photon production with the thicker neutron-skin is less than that with the thinner neutron-skin especially for those of higher energy photons. This is because the larger neutron densities inside the thicker neutron-skin get less photon emission through incoherent proton-neutron bremsstrahlung $p+n\rightarrow p+n+\gamma$ processes in the peripheral collisions of $^{208}Pb+^{208}Pb$, especially for emitting higher energy photons. However, effects of proton-skin in $k$ space on the hard photon emission are more pronounced than that of neutron-skin in $r$ space. We can see that it gets more emissions of hard photons when considering the proton skin effect (mainly caused by the surface term in the nucleon kinetic-energy density) than without considering it except in the hard photon energy $E_{\gamma}=250$ MeV. Because the surface term induces larger local momentum for proton than that for neutron in surface area of nucleus, which will increase the emission probability of hard photons through $p+n\rightarrow p+n+\gamma$ processes. The hard photon with energy $E_{\gamma}=250$ MeV may originate from multiple scattering of nucleons below Fermi momentum rather than from collisions of nucleons at HMT. It is not affected by the surface terms of the nucleon kinetic-energy density in which the short-range correlations are included.

\begin{figure}[t!]
\centering
\setlength{\abovecaptionskip}{-6.cm}
\includegraphics[width=0.99\textwidth]{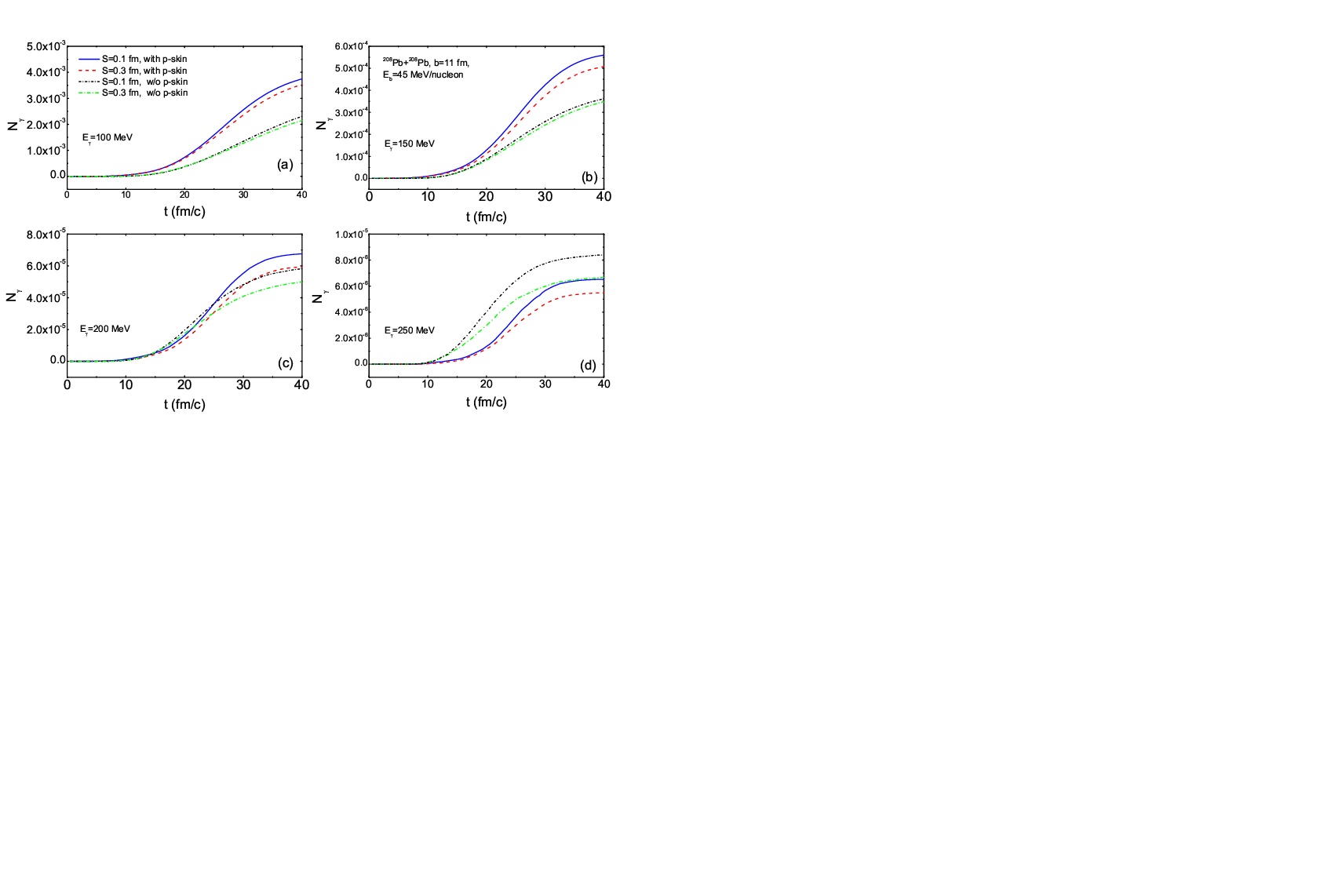}
\caption{Effects of the neutron skin in $r$ space and proton skin in $k$ space on multiplicity of hard photons with energies of $E_{\gamma}=100,\ 150,\ 200$ and $250$ MeV respectively in peripheral collisions $^{208}Pb+^{208}Pb$ at a beam energy of 45 MeV/nucleon.} \label{Multi-t}
\end{figure}

Shown in Fig.~4 are angular distributions of the single differential probability of hard photons with energies of $E_{\gamma}=100,\ 150,\ 200,$ and $250$ MeV respectively in the c.m. frame of the colliding nuclei. We can see that the peak around $\theta=90^{\circ}$ for the angular distribution of hard photons is reproduced as shown in Ref.~\cite{GWM21}. Here, effects of the neutron skin in $r$ space and proton skin in $k$ space on the angular distribution of hard photons in peripheral collisions $^{208}Pb+^{208}Pb$ at a beam energy of 45 MeV/nucleon are compared. With the increasing of hard photon energies, the effect of neutron skin on the production of hard photons increases. Apparently, the thinner neutron skin creates more collision opportunities for neutrons and protons from $p+n\rightarrow p+n+\gamma$ processes in the peripheral collisions, which will make more energetic hard photons. Moreover, the effect of neutron skin on emission of hard photons becomes larger when considering the proton skin in $k$ space. This is because the increasing local momentum of protons near the surface of nucleus makes more colliding for protons with neutrons with neutron skin of $S=0.3$ fm compared with neutron skin of $S=0.1$ fm in the heavy-ion reactions.
As obtained in Fig.~3, effects of proton-skin in $k$ space on the angular distribution of hard photons are also more pronounced than that of neutron-skin in $r$ space. Thus, it will strongly interfere the determination of the size of neutron skin.

\begin{figure}[t!]
\centering
\setlength{\abovecaptionskip}{-5.cm}
\includegraphics[width=0.99\textwidth]{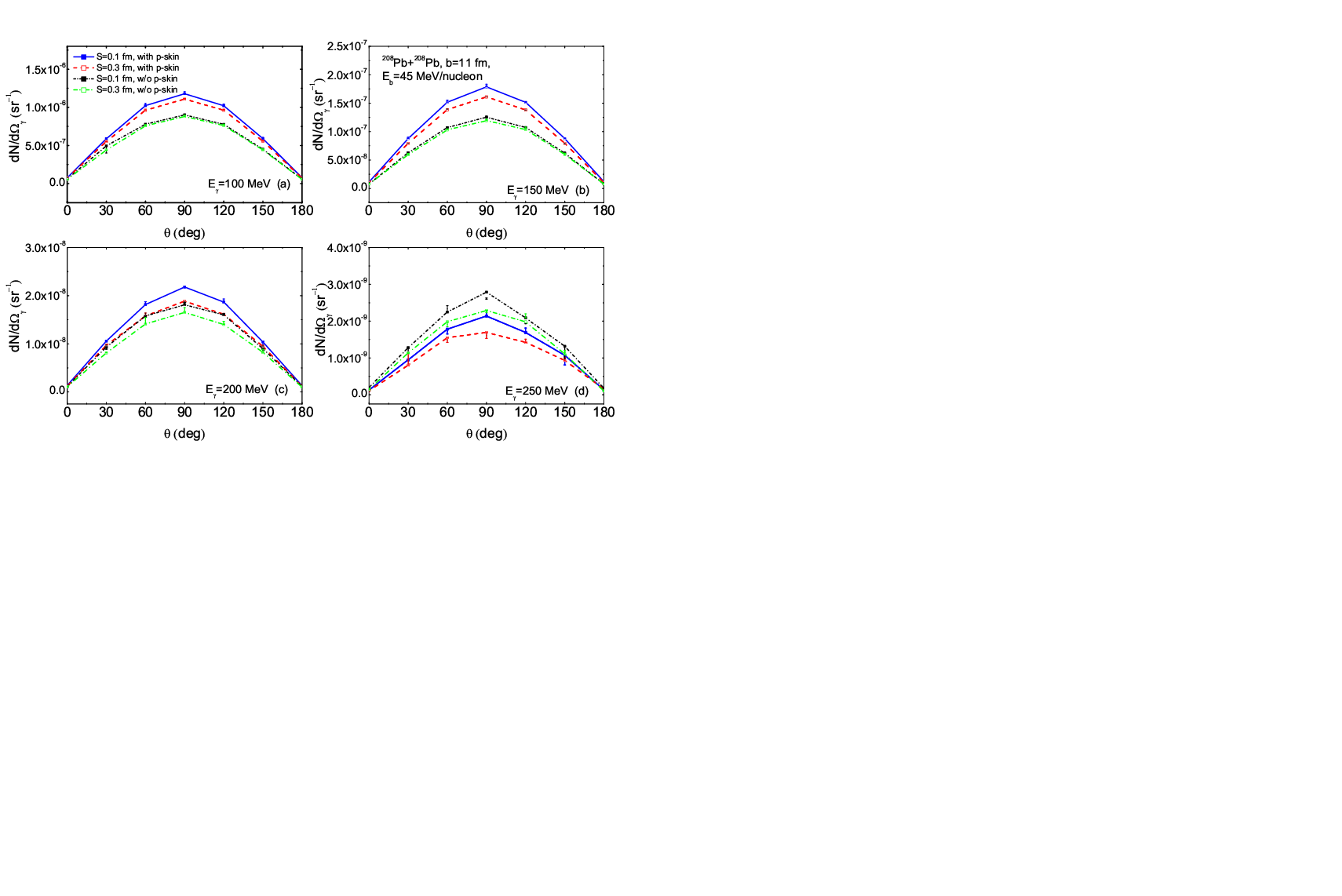}
\caption{Effects of the neutron skin in $r$ space and proton skin in $k$ space on the angular distributions of photons with energies of $E_{\gamma}=100,\ 150,\ 200,$ and $250$ MeV in the c.m. frame of the colliding nuclei from peripheral $^{208}Pb+^{208}Pb$ collisions at $E_b=45$ MeV/nucleon.} \label{Angular}
\end{figure}

\begin{figure}[t!]
\centering
\setlength{\abovecaptionskip}{-3.cm}
\includegraphics[width=0.999\textwidth]{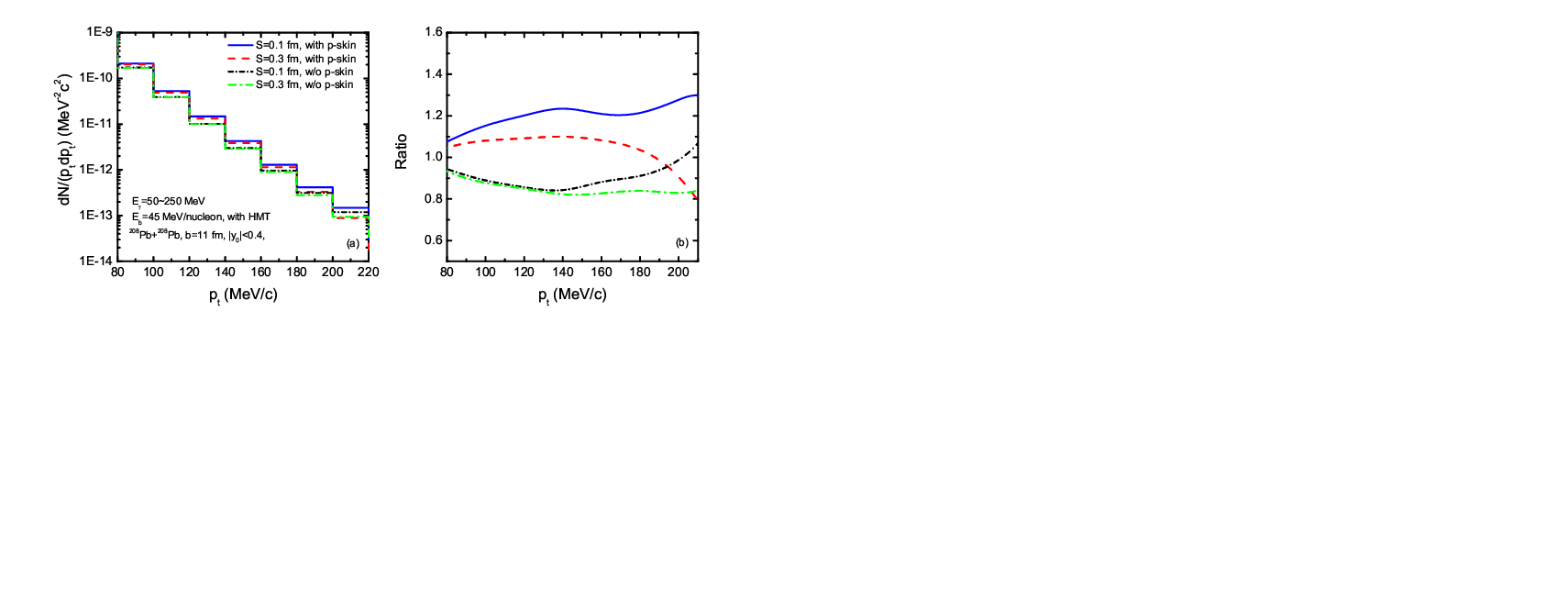}
\caption{Effects of the neutron skin in $r$ space and proton skin in $k$ space on the transverse momentum dependence of hard photons (a) and ratios of each transverse momentum spectrum of hard photons and average value of the transverse momentum spectra for four cases (b) in the midrapidity region of $-0.4\leq y_0\leq0.4$ in peripheral $^{208}Pb+^{208}Pb$ collisions at $E_b=45$ MeV/nucleon.} \label{Trans-M}
\end{figure}

Fig.~5 (a) describes the transverse momentum distributions of the hard photons spectra, $dN/p_tdp_t$ versus $p_t$ ($p_t=\sqrt{p^2_x+p^2_y}$) in the midrapidity region of $-0.4\leq y_0\leq0.4$ ($y_0=y/y_{beam}$ is the ratio of particle rapidity y over beam rapidity $y_{beam}$ in the c.m. frame of colliding nuclei), for peripheral collisions of $^{208}Pb+^{208}Pb$ at a beam energy of $E_b=45$ MeV/nucleon with different neutron skins in $r$ space and with or without proton skins in $k$ space.
It is seen that the spectra show typical exponential shapes considering different neutron skins and proton skins.
Interestingly, when considering the proton skin in $k$ space, effects of the different neutron skins on the transverse momentum spectrum are more obvious compared to without considering it especially at the high transverse momentum. This is the same with the conclusion shown in Fig.~3 and 4.
To evaluate the specific effects of both neutron skin in $r$ space and proton skin in $k$ space on the transverse momentum spectrum of hard photons, shown in Fig.~5 (b) are the ratios of each transverse momentum spectrum of hard photons in Fig.~5 (a) over the average value of the transverse momentum spectra for four cases. We can clearly see that the effect of neutron skin on the above ratio of transverse momentum spectra of hard photons becomes larger with the increase of the transverse momentum. Moreover, from the mentioned ratios of transverse momentum spectra of hard photons for different neutron skins and proton skins, we found that it is more sensitive to the proton skin in $k$ space than to the neutron skin in $r$ space. Thus, the effect of proton skins in $k$ space is not a negligible factor, when determine the size of neutron skin by the emission of hard photons.
Interestingly, the effect of proton skin on the ratio of transverse momentum spectra of hard photons is more obvious than that on the angular distributions of hard photons, making the former a promising observable for investigating the surface properties of the nucleus.

\section{Summary}

Motivated by the new efforts in better understanding/measuring the neutron skin thickness of nuclei, as well as the strong interests of some experimental groups to measure hard photons \cite{Pcom}, we studied the interplay of effects of neutron skins in coordinate space and proton skins in momentum space on hard photons emission in neutron-rich reaction $^{208}Pb+^{208}Pb$ around Fermi energy.
In this work, we adopt the extended Thomas-Fermi approximation for the nucleon kinetic-energy density, in which the SRC effects were incorporated, in the IBUU transport model. Here, the coexistence phenomenon of the proton skins in $k$ space and the neutron skin in $r$ space exists as shown in Ref.~\cite{Cai16}. It is found that the hard photon production with thinner neutron-skin ($S=0.1$ fm) is larger than that with thicker neutron-skin ($S=0.1$ fm) in peripheral collisions especially for those of higher energy photons. With increasing of the hard photon energies, the effect of neutron skins on the emission of the hard photons increases.
Therefore, the production of hard photons from peripheral heavy-ion collisions is a good probe to investigate the size of neutron skin with its characteristics unaltered by final state interactions. Moreover, it is noticed that the effect of neutron skin in $r$ space on the emission of hard photons becomes larger when considering the surface term in the kinetic energy density, leading to the appearance of proton skin in $k$ space.
It reminds us to consider effects of proton skins, when determining the size of neutron skins by the production of hard photons. In addition, the angular distributions of the single differential probability
and the transverse momentum spectrum for hard photons in peripheral collisions of $^{208}Pb+^{208}Pb$ at a beam energy 45 MeV/nucleon were studied in this work. It will provide a basis for the ongoing and planned experiments using hard photons to explore the initial stage in intermediate energy heavy-ion collisions.

Since the surface term in the kinetic energy density of nucleon will affect the surface properties of nuclei, such as the neutron skin of some neutron rich nuclei, it will inevitably have a significant impact on the extraction of proton skin thickness of neutron-deficient isotopes from isotopes near the $\beta$-stability line, which plays an important role in the study of nuclear structure and deserves further studies.

\begin{acknowledgments}
The work is partially supported by the Key Program of Innovation and Entrepreneurship Support Plan for Returned Talents in Anhui Province under Grant No.~2020LCX011 and the Institute of energy, Hefei Comprehensive National Science Center under Grant No.~GXXT-2020-004. B. A. Li is supported in part by the U.S. Department of Energy, Office of Science, under Award No.~DE-SC0013702, the CUSTIPEN (China-U.S. Theory Institute for Physics with Exotic Nuclei) under US Department of Energy Grant No.~DE-SC0009971.

\end{acknowledgments}


\end{document}